\documentclass[%
twocolumn,
 amsmath,amssymb,
 aps,
]{revtex4-1}

\usepackage{graphicx}
\usepackage{dcolumn}
\usepackage{bm}
\usepackage{placeins}
\begin{document}


\title{Effect of filter shape on excess noise performance in continuous variable quantum key distribution with Gaussian modulation}

\author{Hou-Man Chin,$^{1,2*}$ Nitin Jain,$^1$ \\ Darko Zibar,$^2$ Tobias Gehring,$^1$ Ulrik L. Andersen$^1$}

\address{$^1$\mbox{Center for Macroscopic Quantum States, bigQ, Department of Physics, Technical University of Denmark,} 2800 Kongens Lyngby, Denmark\\
$^2$\mbox{Department of Photonics Engineering, Technical University of Denmark,} 2800 Kongens Lyngby, Denmark
}

\email{homch@fysik.dtu.dk}

\date{\today}

\begin{abstract}
An attractive implementation for quantum cryptography is the continuous variable variation, as it relies on standard telecommunication components. Modulating the quantum signal using a Gaussian format is attractive since it has been proven to be secure. This work investigates the effect of the roll-off of a root raised cosine pulse shaping and matched filter on the excess noise performance of a Gaussian modulated quantum key distribution system in a simulated back to back configuration. Contrary to intuition, it is found that the roll-off parameter does not significantly impact the performance of the system.
\begin{description}
\item[PACS numbers]
May be entered using the \verb+\pacs{#1}+ command.
\end{description}
\end{abstract}

\pacs{Valid PACS appear here}
\maketitle

\section{Introduction}

Quantum key distribution (QKD) technology is a key component in securing future telecommunications. The most important property of an encryption scheme implementing QKD is its future proof quality. Current public key encryption schemes rely on assumptions of mathematical complexity. An increase in algorithmic decryption efficiency and/or the advent of quantum computers can breach the security provided by such schemes. In particular encrypted communication can be stored today and sensitive information can be decrypted in the future when new technology became available \cite{Mulholland2017}\cite{Scarani2009}. QKD provides mathematically provable future proof encryption schemes regardless of the capability of an attacker (Eve).

Continuous variable quantum key distribution (CV-QKD) is an attractive method to implement QKD since it utilises standard telecommunication equipment, such as lasers, electro-optical modulators and coherent detectors, and has the potential to operate alongside classical optical telecommunications signals in the same optical fibre through wavelength division multiplexing. CV-QKD requires the quantum signal from the transmitter (Alice) to be sent with extremely low signal powers resulting in a received signal-to-noise ratio of one or less at the receiver (Bob). Compensation of phase noise in addition to clock recovery of the quantum signal is extremely challenging. A common approach is to use pilot tone aided schemes \cite{Laudenbach2017}\cite{Kleis2017}. 

Due to the shot noise limited transmission required by CV-QKD, a popular receiver architecture is coherent balanced detection. Instead of using a phase diverse receiver to recover the quadrature components, a single side band modulated signal is transmitted, similar to classical telecommunications schemes \cite{Erkilinc2014}. This is because given the low optical launch power of the quantum signal, the extra attenuation that would be incurred from propagation through an optical hybrid in addition to the penalty for detecting the in phase and quadrature components separately is not desired. Therefore, an optical heterodyne receiver detection scheme where the signal and local oscillator lasers are offset in frequency is attractive. 

This work investigates the optimization of a pilot tone aided phase error compensation for a heterodyne receiver. Such schemes have previously been demonstrated for M-ary phase shift keying (PSK) signals \cite{Kleis2017}\cite{Laudenbach2017}. Here however we examine the performance of a Gaussian modulation format~\cite{Grosshans2002, Weedbrook2012} using simulations, and optimize the root raised cosine filter roll-off parameter for it.

\section{Excess noise}
The performance of CV-QKD systems depends crucially on the excess noise. In security proofs excess noise is usually modelled by an entangling cloner attack where an eavesdropper injects a part of an entangled state into the transmission line enabling her to optimally extract information~\cite{Weedbrook2012}. However, not only a potential eavesdropper but also imperfections in the hardware and signal processing algorithms contribute to the total excess noise budget. Excess noise severely limits the possible transmission distance and thus has to be reduced as much as possible. The excess noise is therefore a very good measure of system performance. While electronic noise of the receiver is usually the dominant noise source, the second most important source is usually introduced by the error in the phase estimation algorithm. 
\begin{figure*}[t]
   \centering
        \includegraphics[width=120mm]{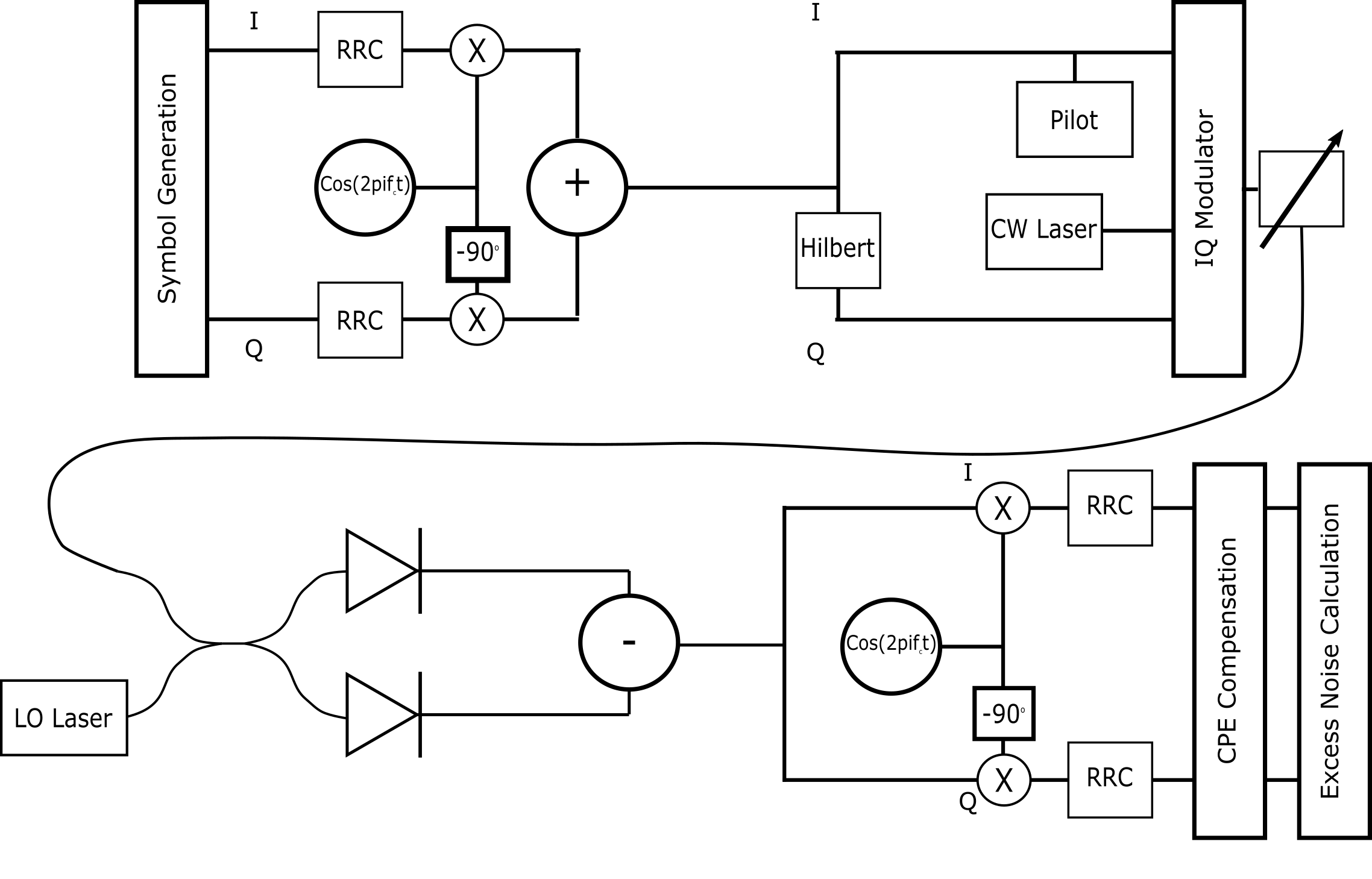}
    \caption{System setup for simulated transmission 50 Mbaud root raised cosine pulse shaped Gaussian modulation quantum signal in a back to back configuration which is received by a coherent heterodyne balanced detector implementing matched filtering}
    \label{fig:systemsetup}
\end{figure*}
Since excess noise has a severe impact on system performance, it must be carefully estimated. For instance it requires a careful calibration of the shot noise of the shot-noise limited receiver. For a Gaussian modulated alphabet, the excess noise in any of the quadratures I and Q, measured by heterodyne detection and expressed in shot noise units (SNU) by normalizing all variances to the shot noise variance $N_{0}$, is given by
\begin{equation}
	\xi = V_{B} - \eta T\left(\frac{V_{A} + 1}{2}\right) - \frac{1}{2} \ ,
\end{equation}
where $V_{A}$ is the modulation variance of Alice's transmitted optical quantum signal, $V_{B}$ is variance estimated from Bob's measurements of the quantum signal at the receiver, $T$ is the transmittance, set to the known trusted loss of the transmission fibre, and $\eta$ is the quantum efficiency of the receiver. In general, the modulation variance of the I and Q quadratures should be the same.


The combination of shot noise $N_{0}$ and electronic (or thermal) noise $N_{det}$ can be determined by measuring the `vacuum state' where there is no input signal to the receiver, i.e. the measured variance is from the local oscillator and thermal noise. $N_{det}$ is then acquired by measuring the variance of the electrical signal when both the local oscillator and the input optical signal to the receiver are disconnected. From this the absolute value $N_{0}$ can be determined. It can be easily imagined that depending on the filter bandwidth or the filter shape used at the receiver, the variance of the shot noise and electronic noise will be affected. Given the sensitivity required for transmission in the quantum regime, this can lead to a significant change in the absolute value of a shot noise unit. Intuitively the excess noise evaluated in this work is a result of the estimation error of the power of the quantum signal caused by an error in the phase compensation.

\section{System setup}
\begin{figure}[htbp]
   \centering
        \includegraphics[width=70mm]{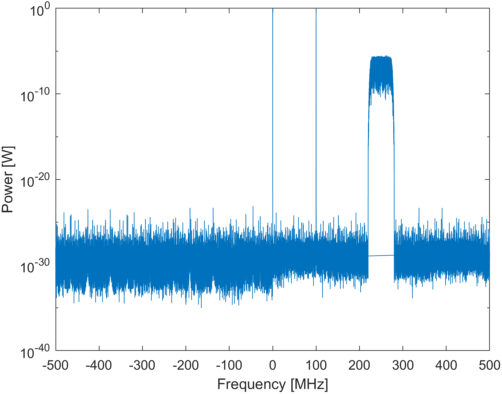}
    \caption{Spectrum of RF signal modulating the transmitter, normalized to pilot power - $\beta = 0.2$}
    \label{fig:SSBSpectrum}
\end{figure}
The system setup used for the simulation is as shown in Fig.~\ref{fig:systemsetup}. A $2^{15}$ long sequence of symbols is generated from a Gaussian distribution for the quantum signal to be modulated at 50 Mbaud. This sequence is upsampled for a simulation bandwidth of 1 GHz. In preparation for single side band (SSB) generation, the upsampled signal is first filtered by a root raised cosine filter and then frequency shifted to 250 MHz. A Hilbert transform is then applied to the signal to generate the SSB modulation. Pilot tones are then inserted at the carrier frequency and at 100 MHz. Note that the pilot tone power is set to be 20 dB greater than that of the quantum signal to ensure that the SNR in the pilot tone recovery is not a limiting factor on the phase correction. Fig.~\ref{fig:SSBSpectrum} shows the spectrum of the signal with RRC $\beta = 0.2$. We simulate this signal modulating a 1 kHz linewidth laser.

A variable optical attenuator adjusts the optical power such that the modulation variance of the quantum signal is within the desired shot noise limited regime (in this work, between 2 and 10 SNU). The optical signal is then fed into the receiver for a back to back architecture. A balanced coherent optical heterodyne receiver is implemented using a 3 dB coupler and two photodiodes with $\eta = 0.85$, $I_{D} = 10^{-9}$. A temperature of 293 K and resistance $R_{L} = 50 \Omega$  is assumed in the simulation for the increased receiver sensitivity in addition to removing the influence of the local oscillator (LO). The LO is another simulated 1 kHz linewidth laser. 

The quantum signal is demodulated using the frequency offset estimation done as per where the pilot tones are low pass filtered using a 50 MHz brickwall filter and the phase error $ \phi (k)$ estimated from them as in \cite{Kleis2017} except that instead for a 8PSK system, the modulation format used here is Gaussian. The quantum signal is down-converted and low pass filtered using a root raised cosine filter matched to theone used at the transmitter. An $\exp (-j \phi (k))$ operation compensates for the phase noise experienced by the quantum signal. The excess noise parameter is then calculated for the quantum signal as described in the previous section.

This simulation was run 305 times per roll-off value to calculate the mean excess noise over $10^{8}$ symbols.

\section{Results}
\begin{figure}[t]
   \centering
        \includegraphics[width=\linewidth]{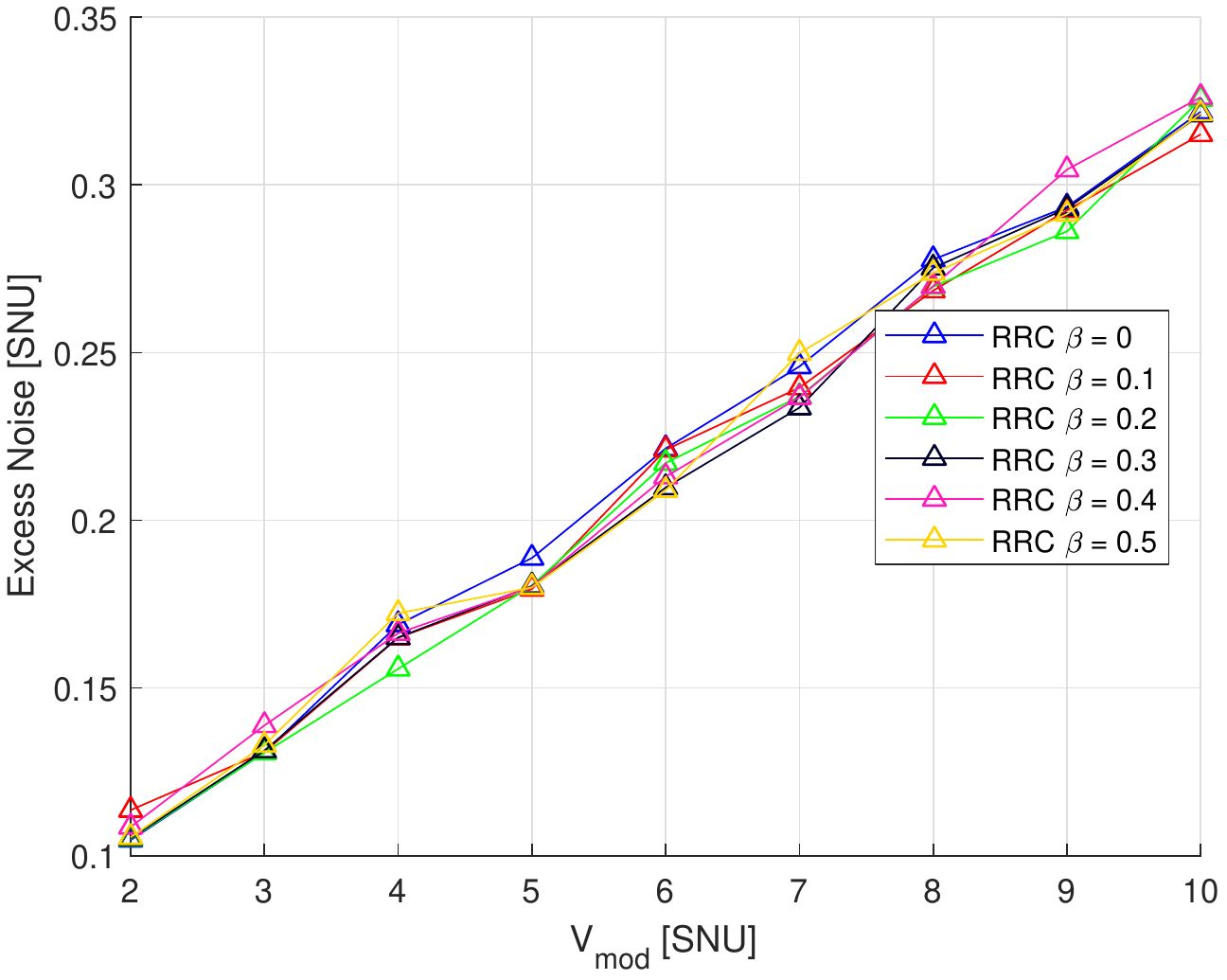}
    \caption{Excess noise for the I quadrature for a range of $V_{mod}$ [SNU] for different root raised cosine filter roll-off values}
    \label{fig:EN_I}
\end{figure}
\begin{figure}[t]
   \centering
        \includegraphics[width=\linewidth]{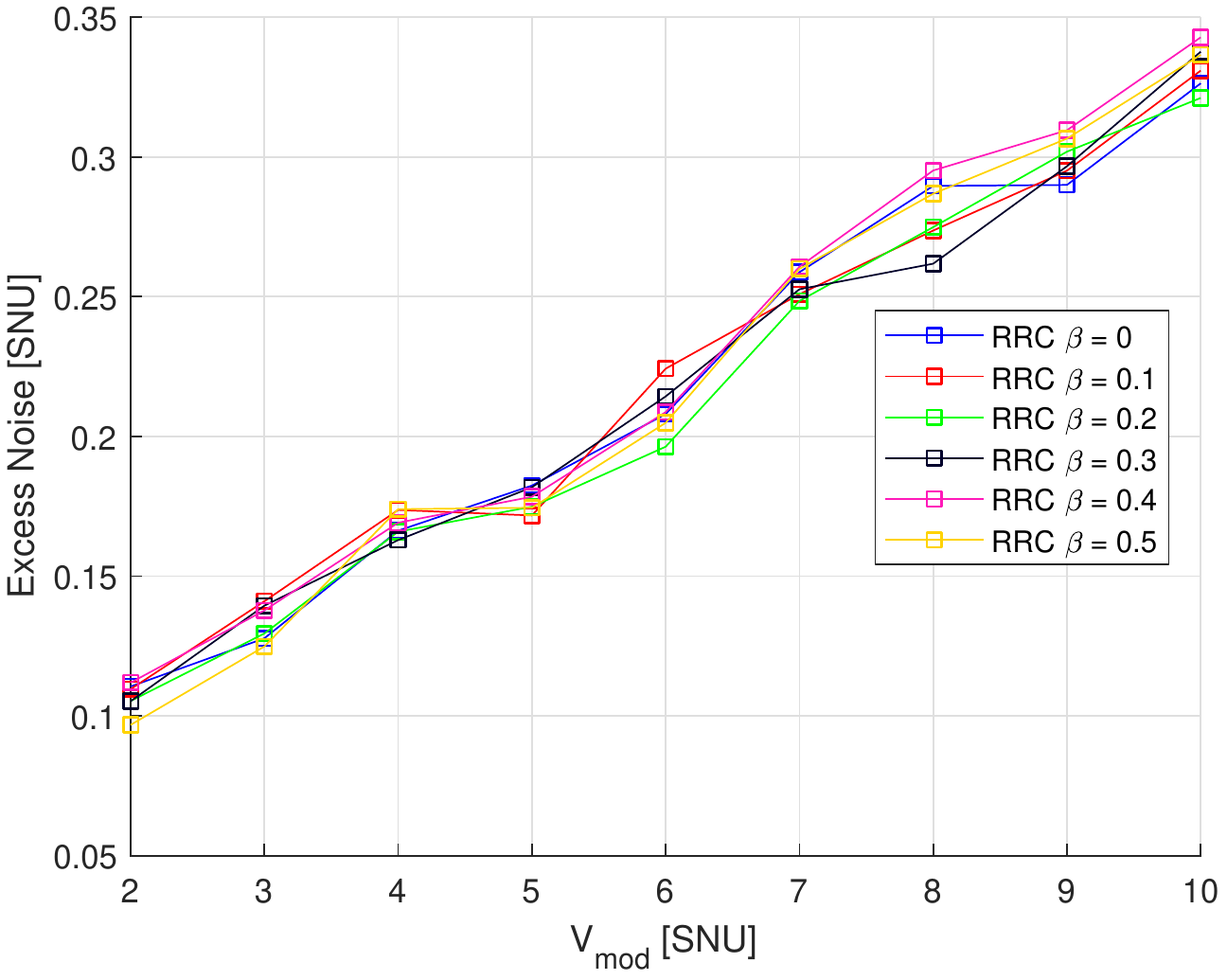}
    \caption{Excess noise for the Q quadrature for a range of $V_{mod}$ [SNU] for different root raised cosine filter roll-off values}
    \label{fig:EN_Q}
\end{figure}
The root raised cosine filter shapes are adjusted using different roll-off factors with 0 being an ideal root raised cosine filter and 0.5 being the largest value used. The shot noise variance is recalculated for each roll-off value and the modulation variances normalized accordingly. 

It can be seen from Fig.~\ref{fig:EN_I} and Fig.~\ref{fig:EN_Q} that there is minimal change in performance of the system with respect to excess noise for both I and Q quadrature. It is however critical that the variance of one shot noise unit is recalculated with respect to the bandwidth filtered at the receiver after frequency down-conversion otherwise a spurious optimization result may be acquired. In the calculation of the excess noise, it should be noted that the transmittance is taken to be 1 due to the back to back nature of the simulation setup i.e. there is no attenuation to transmission through optical fibre. Strictly speaking \textit{T} is not quite equal to 1 even for a back to back system due to the noise at the receiver limiting the co-variance between Alice's transmitted and Bob's measured symbols. The excess noise values displayed here for the naive case of $T = 1$ are therefore slightly higher than would be expected. However for the purposes of this work it is sufficient to show that with respect to the roll-off of matched root raised cosine pulse shaping and filtering, the excess noise performance does not significantly diverge. The excess noise is seen to increase instead with modulation variance. Overall excess noise performance may be improved with further optimization of the low pass filter bandwidth for the pilot signals.

\section{Conclusions}
This work investigated the impact of a range of roll-off values for a root raised cosine pulse shaped/filtered Gaussian modulated QKD system in an optical back to back configuration. There was determined to be no significant impact on the excess noise performance of the system given normalization to the variance of the shot and thermal noise within the bandwidth of the RRC filter. The excess noise increased with respect to the range of modulation variances modulated by Alice at the transmitter linearly. For classical telecommunications engineers this may result in some confusion since the increase in modulation variance would increase the transmitted signal power and therefore the received SNR of the QKD signal which would typically increase signal performance, it may be viewed that the increase in modulation variance provides more power to incur excess noise with.

\subsection*{Acknowledgements}
We also acknowledge funding from Center for Macroscopic Quantum States (bigQ DNRF142) and the Quantum Innovation Center Qubiz.

\bibliography{20180423.bib}

\end{document}